\documentclass[journal=jacsat,manuscript=article]{achemso}
\usepackage[version=3]{mhchem} 
\usepackage[T1]{fontenc}       
\usepackage[usenames, dvipsnames]{color}
\usepackage{longtable}
\usepackage{multirow}
\usepackage{makecell}

\author{Vo Khuong Dien}
\affiliation[National Cheng Kung University]{Department of Physics, National Cheng Kung University, Taiwan}
\author{Nguyen Thi Han}
\author{Thi Dieu Hien Nguyen}{\tiny }
\affiliation[National Cheng Kung University]{Department of Physics, National Cheng Kung University, Taiwan}
\affiliation[Thai Nguyen University of Education]{Department of Chemistry, Thai Nguyen University of Education, Viet Nam}
\author{Thi My Duyen Huynh}
\author{Hai Duong Pham}
\affiliation[National Cheng Kung University]{Department of Physics, National Cheng Kung University, Taiwan}
\email{phamduong477@gmail.com}
\author{Ming Fa-Lin}
\email{mflin@mail.ncku.edu.tw}
\affiliation[National Cheng Kung University]{Department of Physics, National Cheng Kung University, Taiwan}
\affiliation[National Cheng Kung University]{Hierarchical Green Energy Materials, Hi- esearch Center, National Cheng Kung University, Taiwan}

\title{Geometric and Electronic Properties of Li\textbf{$_2$}GeO\textbf{$_3$}}
\begin{document}
\begin{abstract}
The 3D ternary Li$_2$GeO$_3$ compound, which could serve as the electrolyte material in   Li$^+$-based batteries, exhibits an unusual lattice symmetry (orthorhombic crystal), band structure, charge density distribution and density of states. The essential properties are fully explored through the first-principles method. In the delicate calculations and analyses, the main features of atom-dominated electronic energy spectrum, space-charge distribution, and atom-/orbital-projected density of states are sufficient to identify the critical multi-orbital hybridizations of the chemical bonds: 2s-(2p$_x$, 2p$_y$, 2p$_z$) and (4s, 4p$_x$, 4p$_y$, 4p$_z$)-(2s, 2p$_x$, 2p$_y$, 2p$_z$), respectively, for Li-O and Ge-O. This system possesses a large indirect gap of Eg$=$3.77 eV. There exist a lot of significant covalent bonds, with an obvious non-uniformity and anisotropy. In addition, spin-dependent magnetic configurations are completely absent. The theoretical framework could be developed to investigate the important features of anode and cathode materials related to lithium oxide compounds. 
\end{abstract}
\section{INTRODUCTION}
\par\noindent

Increasing demands for storage of electricity from solar and wind energy, mobile electronic devices, electric vehicles promote the development of cost-effective and reliable electrical energy storage \cite{1,2,3}. Among a large number of possible energy storage technologies such as nickel-cadmium batteries, zinc-manganese batteries, nickel-hydrogen batteries, fuel cells, lead-acid batteries, redox flow batteries, Lithium-ion batteries, etc., the rechargeable Lithium-ion batteries (LIBs) has attracted a great deal of attention  due to their high specific energy, wide working temperature range, high operational voltage and a long cycle life \cite{3}.

A commercial LIB is composed of a negative (cathode) and positive (anode) electrode separated by an electrolyte \cite{3}, in which each component should have a good physical, chemical and material properties, especially for the latter ones to ensure the rapid lithium-ion transmission, is compatible with the electrodes and chemically inert at the same time. Apparently, a drastic change of geometric structures is revealed in the cathode/electrolyte/anode materials. The structural transformations between two-metastable configurations during the battery operation is rather complex and thus very difficult to solve.

In general, the cathode and anode systems of Li$^+$-based batteries belong to a class of solid-state materials, such as the three-dimensional (3D) ternary LiFe/Co/NiO \cite{4,5,6} and Li$_4$Ti$_5$O$_{12}$/graphite compounds \cite{7,8}, respectively, whereas conventional electrolytes belong to liquid states, which come with have potential security risks concerning volatilization, flammability and explosion. Recent experimental studies indicated that secondary batteries using inorganic solid electrolytes would be the ultimate batteries to resolve the safety issues \cite{9}. Moreover, the battery cell design would be simplified with solid electrolytes \cite{10}. Some candidate systems for solid-state electrolytes are Li$_3$OCl \cite{11}, Li$_2$SiO$_3$ \cite{12,13} and Li$_2$GeO$_3$ \cite{14,15,16,17,18}. Among them, the later one has attracted significant research interest due to the following reasons: the 3D ternary Li$_2$GeO$_3$ is produced by relatively simple and direct methods and possesses a reliable ionic conductivity (1.5$\times$10$^{-5}$ ($\Omega$.cm)$^{-1}$) \cite{15}, which is promising as an alternative for the conventional liquid electrolytes. The ternary Li$_2$GeO$_3$ compound was also reported as a new Li$^+$ superionic conductor exhibiting an excellent electrochemical performance, such as cycle stability, charge capacity (725 mAhg$^{-1}$) and rate capability (810 mAhg$^{-1}$ after 35 cycles)\cite{18}. Furthermore, other features, such as the large pyroelectric constants (5 times larger than that of tourmaline), high piezoelectric constant (5 times larger than that of $\alpha$-quartz), transparency in the visible region and moderate mechanical impedance, can be used in various pyroelectric, piezoelectric, and acousto-optic devices \cite{19,20}.

On the theoretical side, a number of studies using the molecular dynamics (MD) and density functional theory (DFT) were investigated on the phase stability \cite{21,22,23}, defect structure \cite{24,25}, electronic properties \cite{23} and the Li$^+$ transport mechanisms \cite{22,24,25,26} of the electrodes or electrolyte materials of LIBs. The complicated atomic interactions (multi-orbital hybridizations) are generated from the diversified physical or chemical properties and the complex material structure. Therefore, they significantly affect the parameters that control ion-transport in the solid-state LIBs \cite{27}. However, systematic investigations into the interaction of the chemical bonds of Li$^+$-rich 3D ternary compounds as a practical electrolyte for LIBs are still rather limited. Especially, the multi-orbital hybridization that is related to the essential properties of the Li$_2$GeO$_3$ compound is absent in investigations up to now.
 
The previous numerical studies based on VASP simulations are sufficient in developing the theoretical framework for understanding the diversified material/physical/chemical phenomena. This framework has been successfully used to conduct systematic investigations of one-dimensional (1D) graphene nanoribbons \cite{28}, two-dimensional (2D) graphene/silicene with chemical modifications \cite{29,30} and the three-dimensional (3D) ternary Li$_4$Ti$_5$O$_{12}$ compound \cite{7}. Through the delicate analysis, the diversified phenomena of the geometric, electronic and magnetic properties due to different dimensionalities, planar or buckled honeycomb lattices, layer numbers, stacking configurations, adatom chemisorptions, guest-atom substitutions and bulk properties of 3D materials can be fully understood. Therefore, this calculation might be very suitable for investigating the extraordinary properties in a lot of complex oxide compounds, eg., Li$_2$SiO$_3$, Li$_2$GeO$_3$, Li$_4$Ti$_5$O$_{12}$, LiFe/Co/NiO, in main-stream Li$^+$-based batteries.     

In this paper, the geometric symmetries and electronic properties of the 3D ternary Li$_2$GeO$_3$ compound (the electrolyte material of Li$^+$-based batteries) are systematically investigated. The state-of-the-art analysis conducted on the various chemical bonds in a large unit cell, the band structure with atomic domination, the atom-/orbital projected density of states (DOS) and the spatial charge density is capable of providing the critical multi-orbital hybridizations. The spin density distribution, the spin-degenerate/spin-split energy bands around the low-energy regions, and the net magnetic moment also be examined in the detail as to whether the magnetism could exist in this compound. The theoretical predictions on the relaxation structure, the valence states, the whole energy spectrum and the band gap could be examined via Powder X-ray Diffraction (PXRD)/Tunneling Electron Microscopy (TEM)/Scanning Electron Microscopy (SEM)/Scanning Tunneling Microscopy (STM), Angle-Resolved Photo Emission Spectroscopy (ARPESS), Scanning Tunneling Spectroscopy (STS) and optical absorption spectra, respectively. In addition, whether a combination of the phenomenological models with the numerical simulations can be achieved in a reliable manner to clarify independent cases is very important, since this linking is for a full exploration of various properties, such as optical, magneto-electronic and transport properties. Therefore, the close relationship between the tight-binding model and the first-principle calculations is thoroughly examined. The present work provides more perceptive insights into the understanding of the diversified chemical bonding, as well as the electronic properties of Li$_2$GeO$_3$ for the future promising electrolytes of LIBs.

\section{2. COMPUTATIONAL DETAILS}
\par\noindent

We used the density functional theory method via the Vienna Ab-initio Simmulation Package (VASP)\cite{31} to perform the optimization of crystal structures and the calculation of the electronic properties. The Perdew-Burke-Ernzerhof (PBE) generalized gradient approximation was used for the exchange-correlation functional\cite{32}. The interaction between the valence electrons and ions core was evaluated by the projector augmented wave (PAW) method \cite{33}. The cutoff energy for the expansion of the plane wave basis set is 600 eV for all calculation. The Brillouin zone was integrated with a special k-point mesh of 9$\times$9$\times$9 and 21$\times$21$\times$21 in the Monkhorst-Pack sampling technique \cite{34} for the geometric relaxation and electronic calculation, respectively. The convergence condition of the ground-state is set to be 10$^{-6}$ eV between two consecutive simulation steps, and all atoms were allowed to fully relax during the geometry optimization until the Hellmann-Feynman force acting on each atom was smaller than 0.01 eV/{\AA}. Spin-polarized calculations were performed for the geometry optimization and the calculation of the band structure. In the k-point sample, the cutoff energy has been checked for convergence of the calculations. 
\section{3. RESULTS AND DISCUSSIONS}
\subsection{3.1 Optimized structure}
    \begin{figure}[htb]
                \includegraphics[scale=0.15]{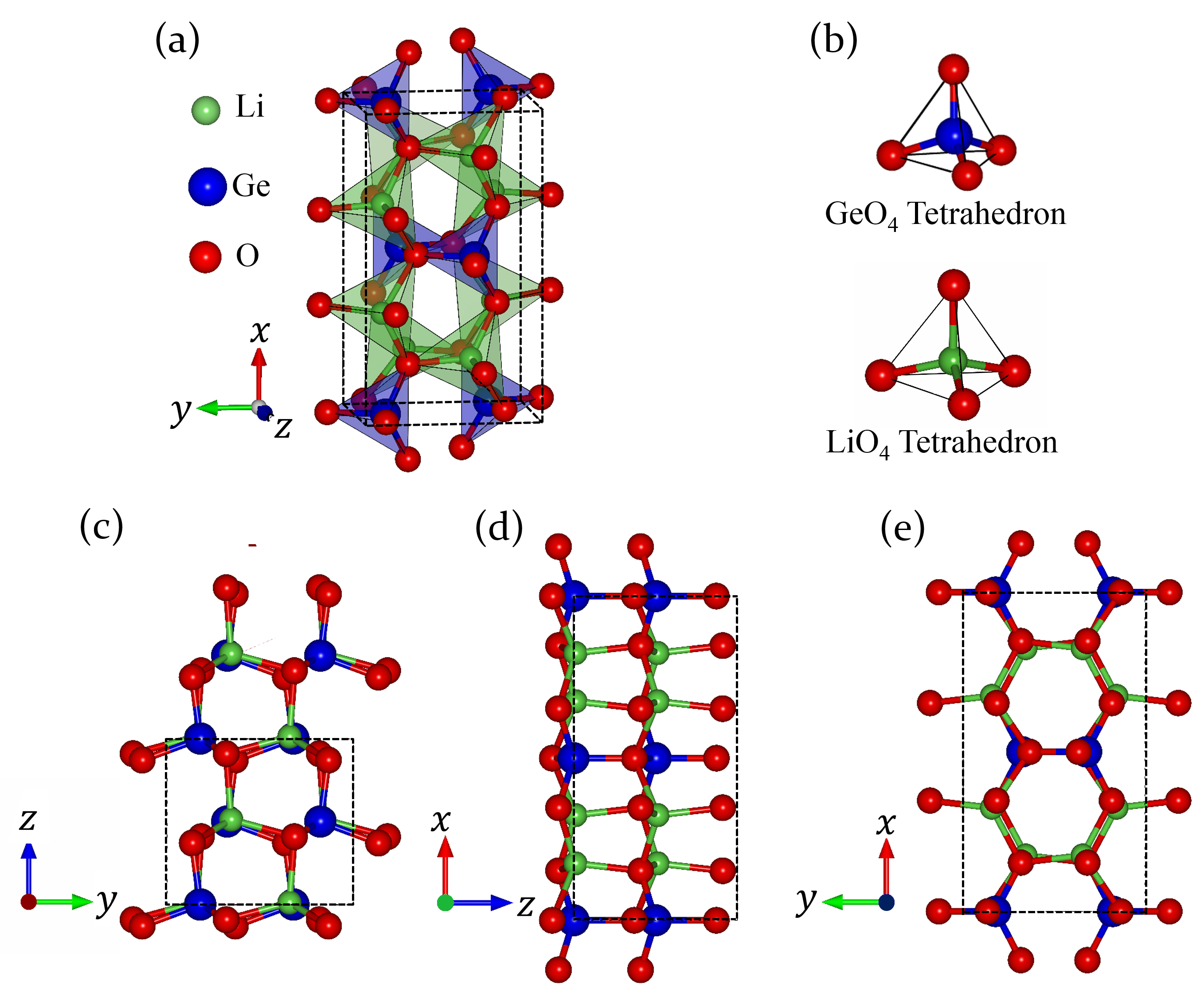}
                \caption{(a) Polyhedron structure of Li$_2$GeO$_3$, (b) Oxygen (O) coordination around the Lithium (Li), Germanium (Ge) atoms. The geometric structure of Li$_2$GeO$_3$ along the different projections (c) (100), (d) (010) and (e) (001).}
                \label{fgr:1}
              \end{figure}
   
   In order to clearly illustrate the complex physical, chemical and material environments in the 3D ternary Li$_2$GeO$_3$ compound, one of the meta-stable configurations is chosen. The geometric configuration of Li$_2$GeO$_3$, as shown in Fig. 1, corresponds to the orthorhombic structure with the Cmc21 space group. The conventional cell considered in this work contains 24 atoms (8 Li, 4 Ge and 12 O), with the lattice parameters being 9.612 {\AA}, 5.462 {\AA} and 4.874 {\AA} for the $x$, $y$, and $z$ directions, respectively. The basic structural unit is comprised of corner-sharing [GeO$_4$] and [LiO$_4$] tetrahedra, in which both Li and Ge ions are coordinated by four O ions.
   
   Based on the first-principles calculations for the optimal geometric structure, the 3D ternary Li$_2$GeO$_3$ electrolyte material presents unusual crystal lattice symmetries. Obviously, there exists a highly anisotropic and extremely non-uniform structure. The diverse atomic arrangements are easily observed under projections on different planes. e.g., the geometric structures for the (100), (010) and (001) directions (Figures 1 (c) - 1 (e)), where Li, Ge, and O atoms are represented by the green, blue and red balls, respectively. As shown in Table 1, there exist only two kinds of chemical bonds, the Li-O and Ge-O ones, with the total number of bonds being 32 and 16, respectively. According to the delicate first principle calculations, the Li$_2$GeO$_3$ compound displays modulated bond lengths in the ranges of $\sim$ 1.930 {\AA} - 2.126 {\AA} and $\sim$ 1.725 {\AA} - 1.836 {\AA}; the fluctuation percentages $\Delta$b(\%)$=$ $\left|bond_{max}-bond_{min}\right|$/$bond_{min}$  are over 10.1\% and 6.4\% for the Li-O and Ge-O bonds, respectively. The various chemical bonds, which are generated by multi-orbital hybridizations, will directly reflect the distribution width of the spatial charge density (discussed later). Such complicated behaviors might induce extra theoretical barriers (various hopping integrals) in exploring the phenomenological models. Most importantly, the strong fluctuation of bond lengths, which generate an extremely non-uniform environment, is expected to be very important in achieving the outstanding charging and discharging processes \cite{35}. Due to this, the various intermediate configurations can easily be transformed during the battery operation. 
   
   Up until recently, PXRD was the most efficient technique in identifying the lattice symmetries of 3D materials \cite{36}. Apparently, it is very suitable to directly observe the 3D ternary Li$_2$GeO$_3$ compound. Furthermore, other quantities, such as the particle size and morphology of the sample, can be observed by using SEM \cite{37}. The top view of the nano-materials is observed by using TEM \cite{38}, while the side view is usually tested by using STM \cite{39}. These methods have been successfully utilized to examine the unique geometric structures of graphene-related systems, such as the multi-walled cylindrical structure of carbon nanotubes \cite{40}, folded and scrolled \cite{41,42}, profiles of graphene nanoribbons \cite{43}, as well as stacking configurations and interlayer distances of multi-layer graphene \cite{44,45}. The predicted unusual geometric structure of Li$_2$GeO$_3$, which includes a non-uniform and the highly anisotropic environment, is in good agreement with the available experimental examinations \cite{18,46} (Table 1).
\begin{table}[htb]
 \fontsize{11}{15}\selectfont
			\caption{Structure parameter of 3D ternary Li$_2$GeO$_3$ compound. The experimental results are also listed.}
			\label{tab:my-table}
			\begin{tabular}{c c c c c c c c c c c c}
				\hline\hline
				\multirow{2}{*}{} &
				\multicolumn{3}{c}{Lattice constants} &
				\multirow{5}{*}{} &
				\multicolumn{4}{c}{Atomic coordianates} &
				\multirow{5}{*}{} &
				\multicolumn{2}{c}{Bond length} \\ \cline{2-4} \cline{6-9} \cline{11-12} 
				&
				a (\AA) &
				b (\AA) &
				c (\AA) &
				&
				Li &
				Ge &
				O$_1$ &
				O$_2$ &
				&
				Li-O &
				Ge-O \\ \cline{1-12}
				This work &
				9.612 &
				5.462 &
				4.874 &
				&
				\begin{tabular}[c]{@{}c@{}}(0.175, \\ 0.342,\\ 0.016)\end{tabular} &
				\begin{tabular}[c]{@{}c@{}}(0.0, \\ 0.179,\\ 0.507)\end{tabular} &
				\begin{tabular}[c]{@{}c@{}}(0.0,\\ 0.137,\\ 0.880)\end{tabular} &
				\begin{tabular}[c]{@{}c@{}}(0.153,\\ 0.318,\\ 0.413)\end{tabular} &
				&
				\begin{tabular}[c]{@{}c@{}}1.930 (8)\\ 1.932 (8)\\ 1.951 (8)\\ 2.126 (8)\end{tabular} &
				\begin{tabular}[c]{@{}c@{}}1.725 (8)\\ 1.833 (4)\\ 1.836 (4)\end{tabular} \\ 
				\begin{tabular}[c]{@{}c@{}}X-ray\cite{45}\\ diffraction \end{tabular} &
				9.632 &
				5.479 &
				4.842 &
				&
				- &
				- &
				- &
				- &
				&
				- &
				- \\ 
				\begin{tabular}[c]{@{}c@{}}X-ray\cite{18}\\ diffraction \end{tabular} &
				9.602 &
				5.502 &
				4.849 &
				&
				\begin{tabular}[c]{@{}c@{}}(0.176,\\ 0.344,\\ 0.015)\end{tabular} &
				\begin{tabular}[c]{@{}c@{}}(0.0,\\ 0.178,\\ 0.5)\end{tabular} &
				\begin{tabular}[c]{@{}c@{}}(0,\\ 0.190,\\ 0.879)\end{tabular} &
				\begin{tabular}[c]{@{}c@{}}(0.162,\\ 0.327,\\ 0.419)\end{tabular} &
				&
				- &
				- \\ \hline\hline
			\end{tabular}
		\end{table}
\subsection{3.2 Electronic properties}
      \begin{figure}[htb]
	\includegraphics[scale=1.2]{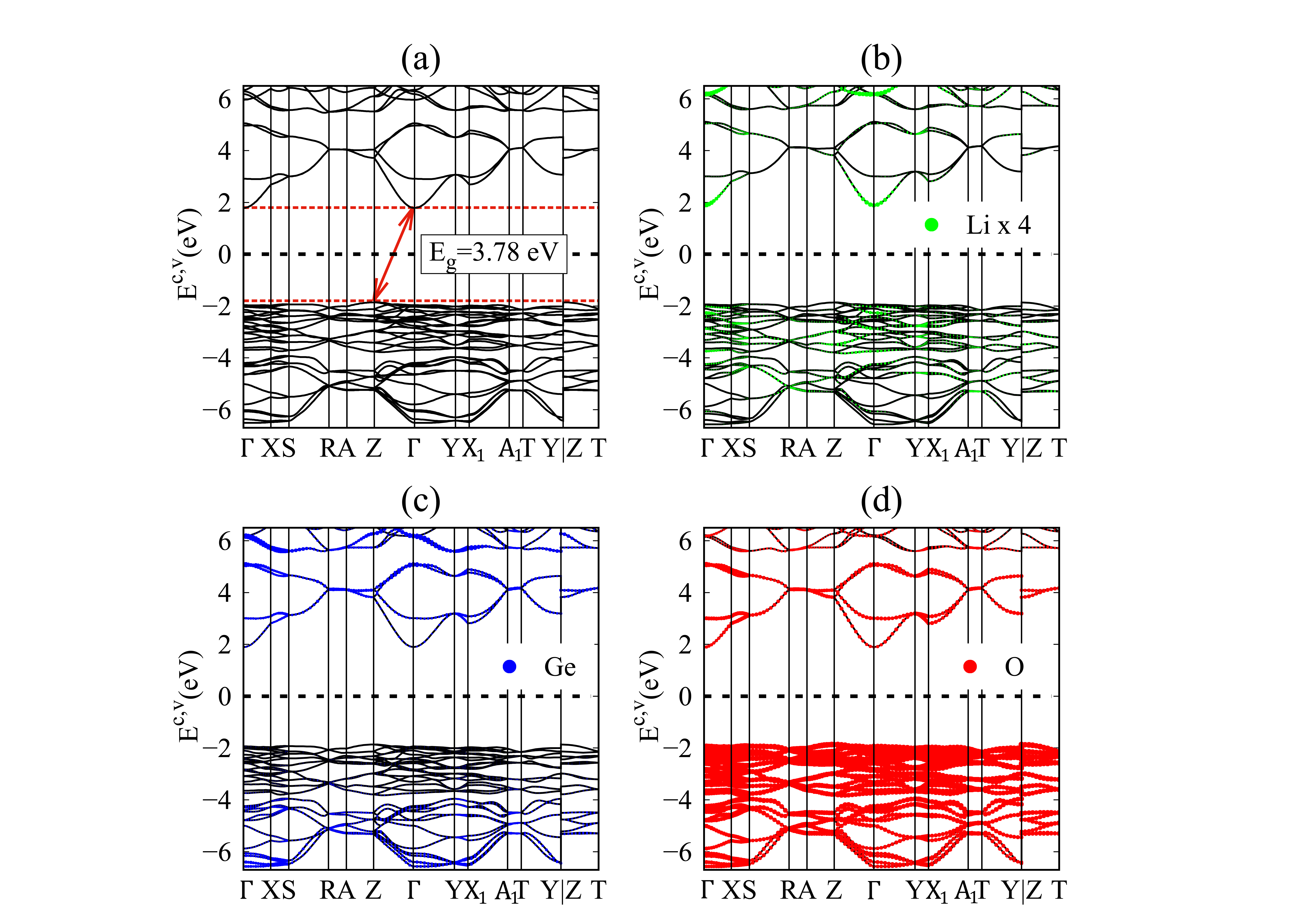}
	\label{fgr:2}
	\caption{(a) Electronic energy spectrum for the Li$_2$GeO$_3$ compound, for the specific (b) Lithium (enlarge four times), (c) Germanium and (d) Oxygen dominances.}
\end{figure}

The 3D ternary Li$_2$GeO$_3$ compound, a Li$^{+}$-based electrolyte material, exhibits unusual geometric structures and thus diversified electronic properties. The calculated band structure of the Li$_2$GeO$_3$ compound is presented in Fig. 2 (a). The zero-energy at the middle of the valence and conduction band is regarded as Fermi energy level. The presence of numerous valence and conduction energy sub-bands in the calculation band structure is a consequence of the many outer orbitals and atoms in the large unit cell. The energy dispersions, which are shown along the high symmetry points, have strongly anisotropic behavior. For example, there exist parabolic, oscillatory and partially flat dispersions. In addition, sub-band anti-crossing, crossing and non-crossing behaviors come into existence frequently, consequently creating the extremely complicated electronic structure. That is to say, it is very difficult to identify/distinguish/examine the various structures for the different sub-bands, as well as to characterize the width of each energy sub-band. Not only that, but plenty of band-edge states might come into existence in the high-symmetry points. These critical points in the energy-wave-vector space would induce the unique van Hove singularities \cite{47} and thus create the strong absorption structures in the optical properties \cite{48}. 

In general, the unoccupied states are highly asymmetric to the occupied states about the Fermi level; particularly, the latter have more energy sub-bands and therefore, are the dominating ones. This behavior might be closely related to the complex orbital hybridizations in the Li-O and Ge-O bonds. Most importantly, the extrema of the conduction band and the valence band are located at the $\Gamma$ and Z point in the first Brillouin zone, respectively, which thus leads to a large indirect band gap of 3.77 eV. This indicates that the Li$_2$GeO$_3$ compound has an excellent electrochemical stability and provides a negligible electronic conductivity\cite{26}. However, the investigations for the band structure of the Li$_2$GeO$_3$ compound are rather limit up to now; therefore, the present results may provide a helpful information for the future work. It is also noted that, under this large band gap, the spin split/degeneracy of the conduction and the valence band near the Fermi level can be not created. Therefore, no magnetic moment exists in this compound, making the spin density distribution meaningless. The mentioned characteristics will be reflected in the density of states. 

In addition to the electronic band structure, the atom-dominated band structures for valence and conduction states are available to understand the critical chemical bondings. In Figs. 2 (b), 2 (c) and 2 (d), the contribution to electronic states of Li, Ge and O atoms are denoted by the green, blue and red balls, respectively. The effective energy range related to the Li-O and Ge-O chemical bonds lies in the range of  -6.5 eV - 6.5 eV. It is very hard for the eye to observe the Li-dominations (small radii in Fig. 2 (b)) because of only single 2s orbital takes parks in the chemical bonding.  In sharp contrast, it is clearly shown that the Ge and O atoms make a significant contribution to the electronic energy spectrum. Especially the latter dominates almost all the occupied and unoccupied bands, and the largest concentration at the top of the valence band (red balls in Fig. 2 (d)) since they are related to the entirely of the chemical bonds, with the (2s, 2p$_x$, 2p$_y$, 2p$_z$)-orbital energies. Although the contributions are different, almost all atoms are present in the whole band structure, which may elucidate the large modulation of the Li-O and Ge-O chemical bond.

From an experimental aspect, the large band gap due to the extremely strong covalent bonding of the 3D ternary Li$_2$GeO$_3$ compound is directly verified by using optical absorption spectroscopy \cite{49}. Apparently, because our system is an indirect gap material, the optical gap should be larger than 3.8 eV as evidenced in previous work \cite{50}. On the other hand, the predictions on the wave-vector-dependent band structures in the occupied states could be examined by ARPES measurements \cite{51}, which has been utilized in many previous experiments to successfully observe the diversified valence bands in emergent graphene-related materials. For example, the monolayer-like and bilayer-like energy dispersions, respectively, at the K and H symmetry points in the stacked graphite \cite{52}. The parabolic/parabolic and linear dispersions in AB-stacked bilayer/trilayer graphene \cite{53,54}, the  Sombrero-shaped and partly flat band in AB-stacked trilayer graphene\cite{55}. The parabolic dispersions in graphene nanoribbons\cite{56}. Up to now, similar measurement on the unusual energy bands in the 3D ternary Li$_2$GeO$_3$ compound is absent. Further examinations are required for the feature-rich and unique energy spectra, including numerous energy subbands, various energy dispersions and the large indirect energy gap. These are useful in understanding the multi-orbital hybridizations of the chemical bonds.  

\subsection{3.3 Charge densities and orbital hybridizations}    
      \begin{figure}[htb]
	\includegraphics[scale=0.3]{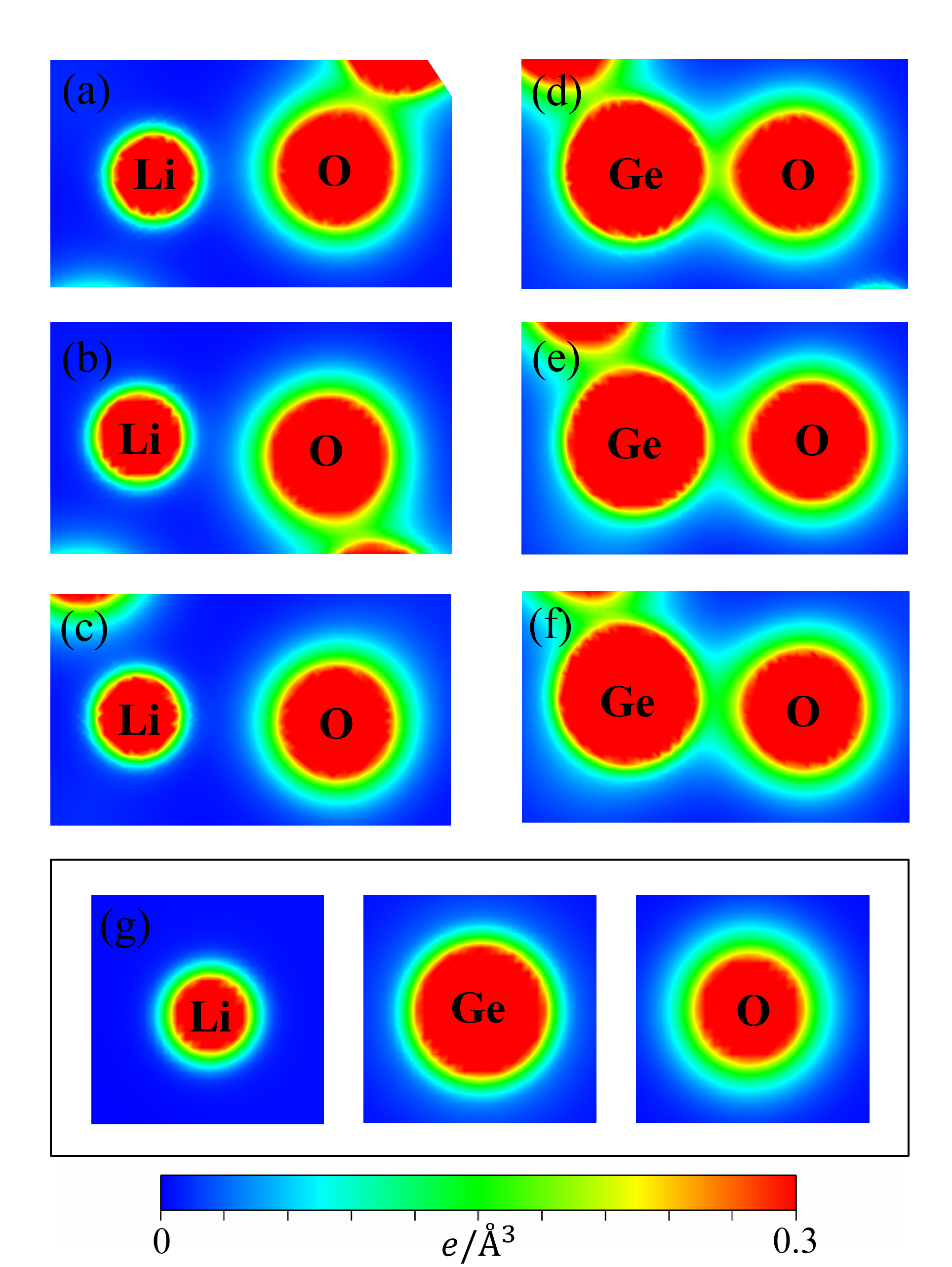}
	\label{fgr:3}
	\caption{The charge density distribution for the shortest/medium/longest (a)/(b)/(c) Li-O bonds and (d)/(e)/(f) Ge-O bonds and (g) for the isolated (Li, Ge, O) atoms.}
\end{figure}
           
The variation of bond lengths can be formed by the complicated contributions of the available orbitals in different atoms. The spatial charge distributions, ($\rho$), as indicated in Figs. 3(a) - 3(g), linking the atom-dominated band structure (Figs. 2(b) - 2(d)) and the atom-/orbital-projected DOS (later discussions on Fig. 5), could provide very useful information on the chemical bondings. 

It can be seen that the carrier densities are strongly dependent on the kind of chemical bonds and very sensitive to the modulation of bond lengths. As for the Li-O bonds (Figs. 3(a) - 3(c)), the diluted charge density around the Li atom is only contributed by the single 2s-orbital. Its effective distribution range is approximately 0.5 {\AA}, as seen by the extension of the yellow to the green parts of the outermost orbitals. Apparently, the two 1s orbitals do not enter into the orbital hybridizations with the O atom,  because it belongs to fully filled electronic configuration. The similar, but wider distribution, which corresponds to the O case, in which the inner and outer region (the heavy red and green parts) is associated with 2s and (2p$_x$, 2p$_y$, 2p$_z$) ones. However, the O-2s orbitals are relatively far away from the Li-atom and remain in almost the perfect spherical shape. Thus, their contributions to the Li-O chemical bonds are negligible. A weak, but significant overlap of distinct orbitals between Li and O atoms is observed. With the increase of the Li-O bond length, this overlap is obviously decreased. As a result, the diverse hopping integrals is presented in the multi-orbital 2s-(2p$_x$, 2p$_y$, 2p$_z$) hybridizations of Li-O bonds. 

In striking contrast, the Ge-O bonds, as clearly shown in Figs. 3 (d) - 3 (f), present very high carrier densities between the Ge and O atoms, indicating a rather stronger chemical bonding than the Li-O bonds and thus, explain why the former is shorter than the latter. Furthermore, we can observe that charge density appears around the Ge atom is larger than that of Li and O ones by reason of the larger atomic number. Its effective region (heavy red, blue-green regions) correspond to the 4s and (4p$_x$, 4p$_y$, 4p$_z$) orbitals. Very interestingly, the deformed spherical distributions of the spatial charge density between Ge-O bonds clearly illustrate that both Ge and O atoms contribute their all valence electrons to form chemical bonds. According to the above-mentioned features, the Ge-O chemical bonds are predicted to exhibit the strong overlap of (4s, 4p$_x$, 4p$_y$, 4p$_z$)-(2s, 2p$_x$, 2p$_y$, 2p$_z$) orbitals. However, with the bond length increase, the modifications on them are observable, as indicated by the decreased carrier density between the Ge and O atoms. The highly non-uniform Li-O and Ge-O chemical bondings might generate numerous difficulties in the phenomenological models (e.g., various hopping integrals).

As mentioned above, the strength of the Ge-O bonds is much stronger than that of the Li-O bonds. The Bader charge analysis of the Li$_2$GeO$_3$ compound has been done to verify this conclusion. It is found that the average effective charge of Li, Ge and O atoms, respectively, are 0.88 e, 2.13 e and -1.3 e. This indicates that the O atom receives the electron from the Li and Ge atoms to forms the more stable configuration. Obviously, the effective charge transfer of Ge attached to the O is much larger than that of the Li bridging to the O and thus, illustrated the former is stronger than the latter ones. It is very important to note that such characteristics are very necessary for the outstanding release and recovery of lithium ions from the electrolyte material during the charging and discharging processes. In other words, the creation of a Li-vacancy is supported by the strong Ge-O bondings and then drives the ion transport.

The density of states (DOS) is characterized as the number of electronic states within a rather small energy range of dE and directly reflects the primary characteristics of the main features of the valence and conduction energy spectra simultaneously. The atom- and orbital-projected DOS, as shown in Fig. 4 and Fig. 5, can be used to fully understand the significant multi-orbital hybridizations of Li-O and Ge-O chemical bonds. As for the Fig. 4, the ternary 3D Li$_2$GeO$_3$ compound exhibits a lot of special structures (Van Hove singularities), including shoulder structures and asymmetric/symmetric peaks. Such features originate mainly from the saddle, local minimum, maximum and dispersionless relation in the energy band structure. Consistently with the electronic band structure, for a wide gap semiconductor of 3.77 eV, the DOS per unit cell of the Li$_2$GeO$_3$ compound vanishes around the centered Fermi level. This disappearance separates the high asymmetry in holes and electrons in the spectra. As can be seen, most of the atoms dominate in the occupied states, but not in the unoccupied ones. The spin-up and spin-down DOS of the former and the latter are the same. In them the difference determines the strength of the net magnetic moment, this indicates an absence of magnetism in the system. Furthermore, the three different kinds of atoms have a significantly different contribution in the conduction and valence bands. For example, the most prominent curve, which belongs to the O atoms, shows a large contribution over the entire energy range, specifically for the top of the valence band. Markedly different, the Li atoms have an almost insignificant contribution, but they are nonetheless very important to create a solid structure of Li$_2$GeO$_3$. Such a result is mainly defined by the number and available valence orbitals of each atom in a unit cell that takes part in the chemical bondings.
      \begin{figure}[htb]
	\includegraphics[scale=1]{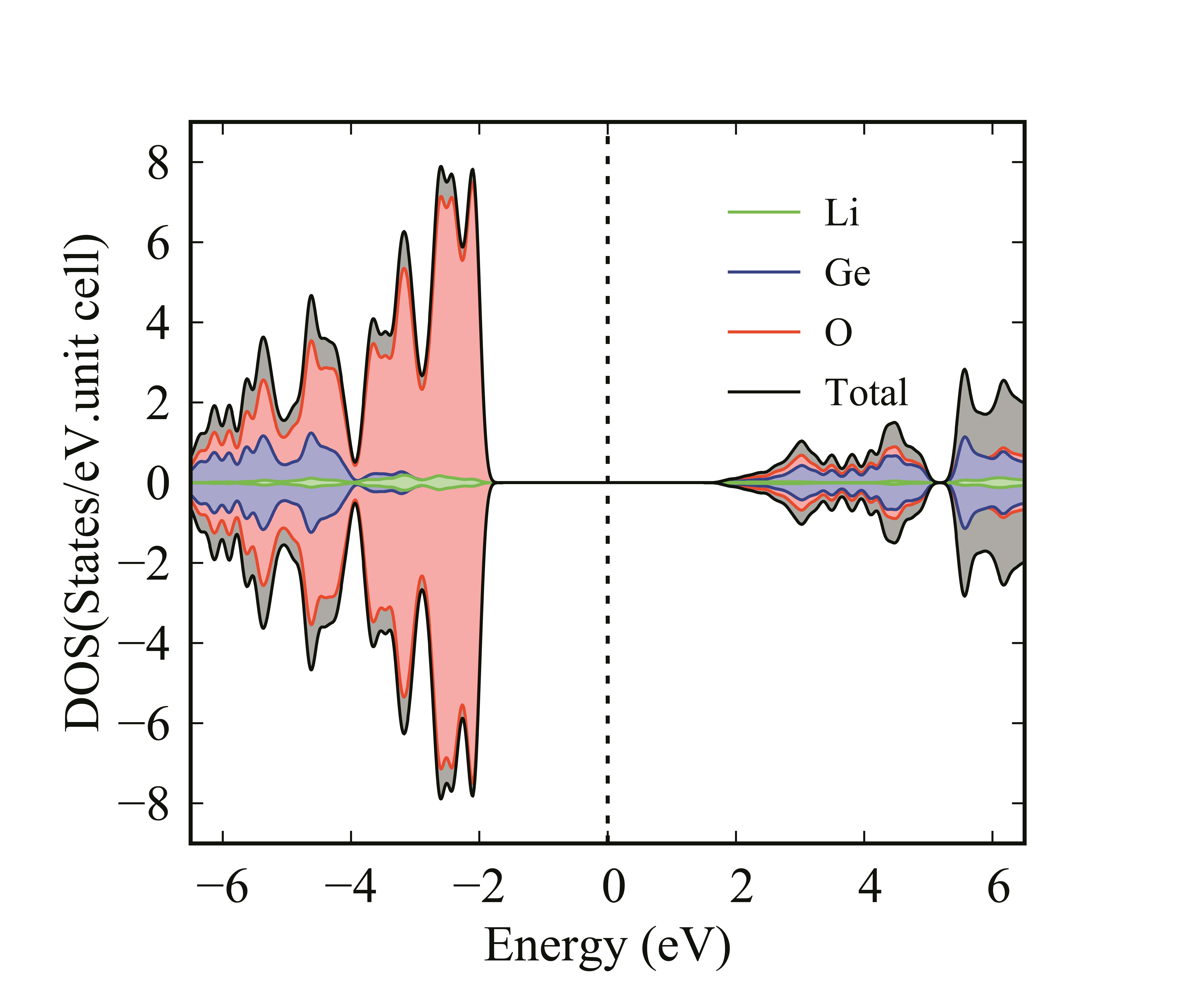}
	\label{fgr:4}
	\caption{The atom-density of states of Li$_2$GeO$_3$ compound.}
\end{figure}

Generally speaking, the various orbitals making an important contribution can be classified into three groups: (I) Li-2s orbital (black curve in the top of Fig. 5), (II) Ge-(4s, 4p$_x$, 4p$_y$, 4p$_z$) orbitals (black, red, green and blue curves in the middle of Fig. 5), and (III) O-(2s, 2p$_x$, 2p$_y$, 2p$_z$) orbitals (black, red, blue and green curves in the lowest of Fig. 5). The effective energy spectrum that relates to the Li-O and Ge-O chemical bonds lie in the range of -6.5 eV - 6.5 eV, as indicated in Fig. 5. The delicate analysis of the enlarged orbital-projected DOS clearly provides the number, energy, intensity and form of Li-2s Van hove singularities. Such structures merge well with the (2p$_x$, 2p$_y$, 2p$_z$)-projected DOS for O atoms, reflecting the importance of the multi-orbital hybridizations, and thus illustrate the large bond length modulation (10.1\%) in the Li-O chemical bonds. However, the contribution the O-2s orbitals into the specific energy range are negligible. Because they are quite low in energy and thus would not combine efficiently with the Li-2s orbitals. Beside the Li-O bonds, the Ge-O bonds also exhibit a very complicated orbital-hybridization. Concerning the Ge atoms, there exist many strong peaks related to the (4s, 4p$_x$, 4p$_y$, 4p$_z$) orbitals. Such peaks play a dominating role in the deeper valence-state region [e.g., E $<$ -3 eV] and the opposite region. Some of them merge with the (2s, 2p$_x$, 2p$_y$, 2p$_z$) peak of the O atoms, clearly illustrating the significance of the Ge-O bonding in the distorted structure. Shortly, the Li-O and Ge-O chemical bonds are deduced to have multi-orbital hybridizations of 2s-(2p$_x$, 2p$_y$, 2p$_z$) and (4s, 4p$_x$, 4p$_y$, 4p$_z$)-(2s, 2p$_x$, 2p$_y$, 2p$_z$), respectively. This unusual characteristic is strongly consistent with the spatial charge density analysis. As far as we can tell, the existence of the complicated multi-orbital hybridization implies that the Li$_2$GeO$_3$ band structure is very difficult to simulate by the tight-binding model.
      \begin{figure}[htb]
	\includegraphics[scale=1]{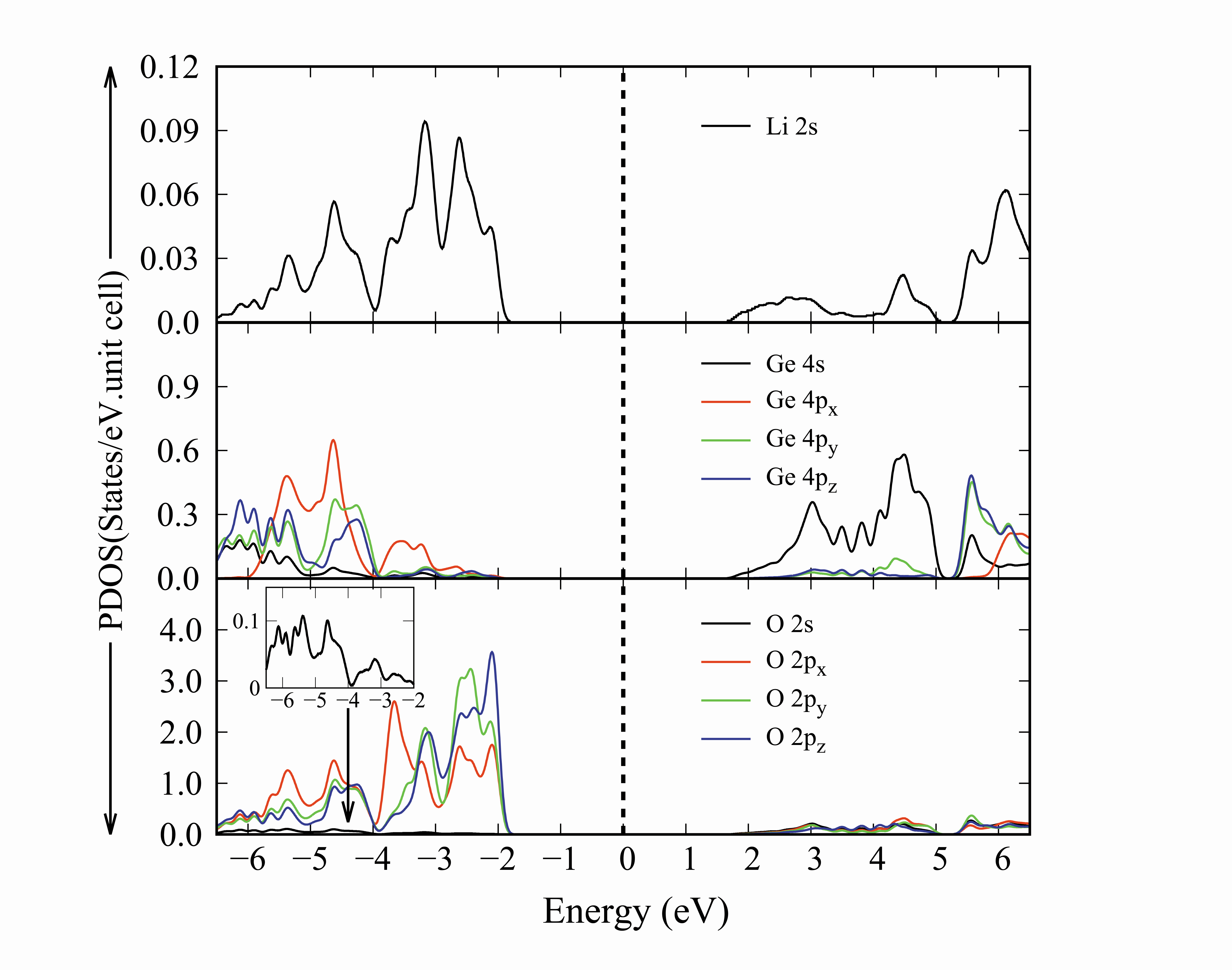}
	\label{fgr:5}
	\caption{The orbital-projected density of states of Li$_2$GeO$_3$ compound.}
\end{figure}

STS measurements \cite{57}, an extension of STM, can be efficiency used to examine the number, form, energy and intensity of special Van-hove singularities in DOS. This powerful method has been successfully identified the diverse electronic properties in GNRs \cite{58,59}, carbon nanotubes, graphene \cite{60}and few-layer graphenes \cite{61,62}. For example, the square-root relation peaks in 1D carbon nanotubes and GNRs \cite{59}, a symmetric V-shape structure for monolayer graphene \cite{61} and finite DOS at Femi level in the AB graphite stacking \cite{60} have all been investigated. The main features of the electronic properties in the 3D ternary Li$_2$GeO$_3$ compound, including the large energy spacing between the occupied and unoccupied bands, the plethora of asymmetric/symmetric peaks, the highly asymmetric electron and hole energy spectra, as well as their widths, could be further investigated with STS. The theoretical predictions of Van Hove singularities, together with the STS measurements are worthy to be identified for the understanding of the complicated multi-orbital hybridizations in the Li-O and Ge-O bonds, and thus the electronic properties of Li$_2$GeO$_3$ compound.

Up to now, LIBs have been quickly developed, owing to their vital importance. Their safety issues can be eliminated by changing the liquid to a solid-state electrolyte. The 3D ternary Li$_2$GeO$_3$ can be adopted to manufacture batteries that are low-cost, durable, and suitable for high-capacity energy storage. Based on the electronic band structure with/without atomic domination, spatial charge distribution, and atom-/orbital-projected DOS calculations, the critical multi-orbital hybridization of Li-O and Ge-O bonds is obtained. Apparently, the Li$_x$Ge$_y$O$_z$ compound has a certain metastable configuration; during battery operation the transformation of many metastable structures (eg. Li$_2$GeO$_3$, Li$_4$GeO$_4$, Li$_2$Ge$_7$O$_{15}$) occurs. Therefore, further systematic studies on other metastable structures are necessary to identify the most prominent evolution paths.

Generally speaking, there exist various chemical bonds in a large unit cell, a complicated energy sub-band with a large band gap, E$_g$=3.77 eV, heterogeneous spatial charge densities in diverse chemical bonds, and multi-orbital hybridizations in atom-/orbital-projected DOS. The first-principles electronic structures might be too complicated to be simulated by the phenomenological models. To be specific, the extremely non-uniform chemical and physical environments, which survive in the large unit cells of the 3D ternary Li$_2$GeO$_3$ compounds, are responsible for the featured electronic structures. Such critical factors cover a lot of multi-orbital hybridizations in the Li-O and Ge-O bondings. The orbital-induced various hopping integral site energies should be included in the significant Hamiltonians simultaneously. Apparently, many parameters will be required to get a good fit. It would be very difficult to create a good Hamiltonian diagonalization, and thus, it is very challenging to achieve a concise physical picture for a full understanding of the featured electronic energy spectra. 

        \section{4. CONCLUDING REMARKS}
          \par\noindent
         The unusual and unique geometric, electronic properties of the 3D ternary Li$_2$GeO$_3$ compound, a potential electrolyte candidate for Li$^+$-based batteries, are investigated by using first principle calculations. The delicate analysis on atom-dominated energy band structure, the charge density distribution in the modulate of the chemical bonds, the atom-/orbital-projected DOS are successfully identified the multi-orbital hybridization in Li-O and Ge-O bonds. This theoretical framework could be generalized for other electrolytic, cathode and anode materials of Li$^+$-based batteries. Furthermore, the extremely non-uniform environment with highly isotropic characteristics in the large unit cell poses a great challenge for the tight-binding model. 
         
         In the current work, the solid-state electrolyte material Li$_2$GeO$_3$, with 24 atoms in a large computational cell is an orthorhombic crystal structure. There exist 32 Li-O and 16 Ge-O chemical bonds, in which each Li/Ge atom is surrounded by four O atoms in the tetrahedral form. Most importantly, the strong covalent bondings create a huge indirect energy gap of E$_g$=3.77 eV. The band structure lying in range -6.5 eV - 6.5 eV is strong relative to the critical chemical bondings. There exist many sub-bands with high anisotropy, strong energy dispersion and frequently anti-crossing/crossing/non-crossing phenomena, and a strong contradistinction of electron and hole states appears near the Fermi level. The band structure with atomic domination, spatial charge density and the atom-/orbital-projected DOS are successfully identify the important multi-orbital 2s-(2p$_x$, 2p$_y$, 2p$_z$) and (4s, 4p$_x$, 4p$_y$, 4p$_z$) - (2s, 2p$_x$, 2p$_y$, 2p$_z$) hybridizations. The theoretical predictions on the structural relaxation are verified by PXRD, TEM and SEM, while the band gap, the occupied electronic states and the special Van-Hove singularities could be examined by optical spectroscopy, ARPES and STS measurements, respectively.
         
         The calculated results clearly illustrate that the 3D ternary Li$_2$GeO$_3$ compound could be used as a solid-state electrolyte. During the battery operation, the structural transformation between the current meta-stable structure with other meta-stable structures are expected to occur at any time. This study is able to provide certain meaningful information about the critical mechanisms in the ion transport. It might be helpful to solve the open issue: the most optimal evolution paths of the Li$^+$ migration during the discharging processes of the all solid-states LIBs.
            \newpage
            
            \par\noindent {\bf Acknowledgments}
            
            This work was supported by the Hierarchical GreenEnergy Materials (Hi-GEM) Research Center, from The Featured
            Areas Research Center Program within the framework of the
            Higher Education Sprout Project by the Ministry of Education
            (MOE) and the Ministry of Science and Technology (MOST 108-
            3017-F-006 -003) in Taiwan.
            \renewcommand{\baselinestretch}{0.2}


\begin{thebibliography}{99}
          	\bibitem{1} Cheng F Liang J, Tao Z, Chen J. Functional materials for rechargeable batteries. \textit{Advanced materials} (2011) \textbf{23}:1695-1715. doi:10.1002/adma.201003587.
          	
            \bibitem{2} Cheng F, Tao Z, Liang J, Chen J. Template-Directed Materials for Rechargeable Lithium-Ion Batteries. \textit{Chemistry of Materials} (2008) \textbf{20}:667-681. doi:10.1021/cm702091q.
            
            \bibitem{3} J.-M. Tarascon and M. Armand. Issues and challenges facing rechargeable lithium batteries. /textit{Nature}. (2001) \textbf{414}:359-367. doi:doi.org/10.1038/35104644.
            
            \bibitem{4} Yoji Sakurai, Hajime Arai, Shigeto Okada, Jun-ichi Yamaki. Low temperature synthesis and electrochemical characteristics of LiFeO2 cathodes.\textit{Journal of Power Sources}. (1997) \textbf{68}:711-715. doi:10.1016/S0378-7753(96)02579-7.
            
            \bibitem{5} Lijun Liu, Zhaoxiang Wang, HongLi, Liquan Chen, Xuejie Huang. Al2O3-coated LiCoO2 as cathode material for lithium ion batteries. \textit{Solid State Ionics}. (2002) \textbf{152-153}:341-346. doi:10.1016/S0167-2738(02)00333-8.
            
            \bibitem{6} Hungru Chen, James A. Dawsona  and  John H. Hardinga. Effects of cationic substitution on structural defects in layered cathode materials LiNiO2. \textit{Journal of Materials Chemistry A}. (2014) \textbf{2}:7988-7996. doi:10.1039/C4TA00637B.
            
            \bibitem{7} Thi Dieu Hien Nguyen, Hai Duong Pham, Shih-Yang Linb  and  Ming-Fa Lin. Featured properties of Li$^+$-based battery anode: Li$_4$Ti$_5$O$_{12}$. \textit{RSC advances}. (2020) \textbf{10}:14071-14079. doi:10.1039/D0RA00818D.
            
            \bibitem{8} Xiang H F, Li Z D, Xie K, Jiang J Z, Chen J J, Lian P C, et al. Graphene sheets as anode materials for Li-ion batteries: preparation, structure, electrochemical properties and mechanism for lithium storage. \textit{Rsc Advances}. (2012) \textbf{2}:6792-6799. doi:10.1039/C2RA20549A.
             
            \bibitem{9} Renjie Chen, Wenjie Qu,  Xing Guo,   Li Li  and  Feng Wu. The pursuit of solid-state electrolytes for lithium batteries: from comprehensive insight to emerging horizons. \textit{Materials Horizons}. (2016) \textbf{3}: 487-516. doi:10.1039/C6MH00218H.
            
            \bibitem{10} Yumei Wang, Shufeng Song, Chaohe Xu, Ning Hu, Janina Molenda, Li Lu. Development of solid-state electrolytes for sodium-ion battery-A short review. \textit{Nano Materials Science} (2019) \textbf{1}:91-100.  doi:10.1016/j.nanoms.2019.02.007.
            
            
            \bibitem{11} Xujie Lu, John W. Howard, Aiping Chen, Jinlong Zhu, Shuai Li, Gang Wu, et al. Antiperovskite Li$_3$OCl Superionic Conductor Films for Solid-State Li-Ion Batteries. \textit{Advanced science}. (2016) \textbf{3}:1500359. doi:10.1002/advs.201500359.
            
            \bibitem{12} Atsuyoshi Nakagawa, Naoaki Kuwata, Yasutaka Matsuda, and Junichi Kawamura. Characterization of Stable Solid Electrolyte Lithium Silicate for Thin Film Lithium Battery. \textit{The Physical Society of Japan}. (2010) \textbf{79}:98-10. doi:10.1143/JPSJS.79SA.98.
            
            \bibitem{13} Yohandys A Zulueta, Minh Tho Nguyen, James A Dawson. Na- and K-Doped Li2SiO3 as an Alternative Solid Electrolyte for Solid-State Lithium Batteries. \textit{The Journal of Physical Chemistry C}. (2020) \textbf{124}: 4982-4988. doi:10.1021/acs.jpcc.9b10003.
            
            \bibitem{14} 
            Shinichi Furusawa, Shun Enokida. Ionic Conductivity of Polycrystalline Li2Ge$_x$Si$_{1-x}$O$_3$ ($x = 0.0 \sim 1.0$). \textit{Key Engineering Materials}. (2011) \textbf{459}:27-31. doi:10.4028/www.scientific.net/kem.459.27.
            
            \bibitem{15} B E Liebert, R A Huggins. Ionic conductivity of Li$_4$GeO$_4$, Li$_2$GeO$_3$ and Li$_2$Ge$_7$O$_{15}$. \textit{Materials Research Bulletin}. (1976) \textbf{11}:533-538. doi:10.1016/0025-5408(76)90235-X.
                    
            \bibitem{16} Guowei Zhao, Kota Suzuki, Masao Yonemura, Masaaki Hirayama, and Ryoji Kanno. Enhancing Fast Lithium Ion Conduction in Li$_4$GeO$_4$-Li$_3$PO$_4$ Solid Electrolytes. \textit{ACS Applied Energy Materials}. (2019) \textbf{2}:6608-6615. doi:10.1021/acsaem.9b01152.
            
            \bibitem{17} Ruijuan Xiao, Hong Li, Liquan Chen. Candidate structures for inorganic lithium solid-state electrolytes identified by high-throughput bond-valence calculations. \textit{Journal of Materiomics}. (2015) \textbf{1}:325-332. doi:10.1016/j.jmat.2015.08.001. 
            
            \bibitem{18} Md Mokhlesur, Rahman Irin, Sultana, Tianyu Yang,  Zhiqiang Chen, Neeraj Sharma,  Alexey M. Glushenkov, Ying Chen. Lithium Germanate (Li$_2$GeO$_3$): A High‐Performance Anode Material for Lithium‐Ion Batteries. \textit{Angewandte Chemie International Edition}. (2016) \textbf{55}:16059-16063. doi:10.1002/anie.201609343.
            
            \bibitem{19} Haussuuhl S, Liebertz J, Stahr S. Single Crystal Growth and Pyroelectric, Dielectric, Piezoelectric, Elastic, and Thermoelastic Properties of Orthorhombic Li$_2$SiO$_3$, Li$_2$GeO$_3$, and Na$_2$GeO$_3$. \textit{Crystal Research and Technology}. (1982) {\textbf{17}}:521-526. doi:10.1002/crat.2170170421.
            
            \bibitem{20} Yu Y S, Prabhu S S, Perkowitz S, Kim S C. Polariton Modes and Materials Parameters in Li$_2$GeO$_3$. \textit{Physics Review B}. (1997) \textbf{56}:5046-5048. doi:10.1103/PhysRevB.56.5046.
            
            \bibitem{21} Yi Zhang, Yusheng Zhao, and Changfeng Chen. Ab initio study of the stabilities of and mechanism of superionic transport in lithium-rich antiperovskites. \textit{Physical review B} (2013) \textbf{87}: 134303. doi:10.1103/PhysRevB.87.134303.
            
            \bibitem{22} Alexandra Emly, Emmanouil Kioupakis, Anton Van der Ven. Phase Stability and Transport Mechanisms in Antiperovskite Li$_3$OCl and Li$_3$OBr Superionic Conductors. \textit{Chemistry of Materials}. (2013) \textbf{25}:4663-4670. doi:10.1021/cm4016222.
            
            \bibitem{23} Farnaz Kaboudvand, Julija Vinckeviciute, Sanjeev Kolli, Maxwell D. Radin, Anton Van der. Phase Stability and Electronic Structure of Tin Sulfide Compounds for Li-ion Batteries. \textit{The journal of Physical chemistry C} (2019) \textbf{123}:29086-29095. doi:10.1021/acs.jpcc.9b06902.
            
            \bibitem{24} N Kuganathan, L H Tsoukalas, A Chroneos. Defects, dopants and Li-ion diffusion in Li$_2$SiO$_3$. \textit{Solid State Ionics}. (2019) \textbf{335}:61-66. doi:10.1016/j.ssi.2019.02.019.
            
            \bibitem{25} Ziheng Lu, Chi Chen, Zarah Medina Baiyee, Xin Chen, Chunming Niub and Francesco Ciuccia. Defect chemistry and lithium transport in Li$_3$OCl anti-perovskite superionic conductor. \textit{Physical Chemistry Chemical Physics}. (2015) \textbf{17}:32547-32555. doi:10.1039/C5CP05722A.
            
            \bibitem{26} Musheng Wu,  Bo Xu, Xueling Lei,  Kelvin Huang  and  Chuying Ouyang. Bulk properties and transport mechanisms of a solid state antiperovskite Li-ion conductor Li$_3$OCl: insights from first principles calculations. \textit{Journal of Materials Chemistry A}. (2018) \textbf{6}:1150-1160. doi:10.1039/C7TA08780B.
            
            \bibitem{27} Qing Zhao, Sanjuna Stalin, Chen-Zi Zhao and Lynden A Archer. Designing solid-state electrolytes for safe, energy-dense batteries. \textit{Nature Reviews Materials}. (2020) \textbf{5}:229-252. doi:10.1038/s41578-019-0165-5.
            
            \bibitem{28} Yu-Tsung Lin, Hsien-Ching Chung, Po-Hua Yang, Shih-Yang Lin and Ming-Fa Lin. Adatom bond-induced geometric and electronic properties of passivated armchair graphene nanoribbons. \textit{Physical Chemistry Chemical Physics}. (2015) \textbf{17}:16545-16552. doi:10.1039/C5CP02226F.
            
            \bibitem{29} Tran, N. T. T, Lin S Y, Lin C Y and Lin M F. Geometric and electronic properties of graphene-related systems: Chemical bonding schemes. \textit{CRC Press}. (2017). doi:10.1201/b22450.
            
            \bibitem{30} Shih-Yang Lin,   Shen-Lin Chang, Ngoc Thanh Thuy Tran, Po-Hua Yang and  Ming-Fa Lin. H-Si bonding-induced unusual electronic properties of silicene: a method to identify hydrogen concentration. \textit{Physical Chemistry Chemical Physics}. (2015) 17:26443-26450. doi:10.1039/C5CP04841A.
            
            \bibitem{31} G. Kresse and J. Hafner. Ab initio molecular dynamics for liquid metals. \textit{Physical review B}. (1993) \textbf{47}:558-561. doi:10.1039/C5CP04841A.
            
            \bibitem{32} John P Perdew, Kieron Burke, and Matthias Ernzerhof. Generalized Gradient Approximation Made Simple. \textit{Physical review Letter}. (1996) \textbf{77}:3865-3868. doi:10.1103/PhysRevLett.77.3865.
            
            \bibitem{33} Kresse G and Joubert D. From ultrasoft pseudopotentials to the projector augmented-wave method. \textit{Physical review B}. (1999) \textbf{59}:1758-1775. doi:10.1103/PhysRevB.59.1758. 
            
            \bibitem{34} Wisesa P, McGill K A and Mueller T. Efficient generation of generalized Monkhorst-Pack grids through the use of informatics. \textit{Physical Review B}. (2016) \textbf{93}:155109. doi:10.1103/PhysRevB.93.155109.
            
            \bibitem{35} Y. Wang, W. Richards, S. Ong, et al. Design principles for solid-state lithium superionic conductors. \textit{Nature Mater}. (2015) \textbf{14}:1026-1031. doi: 10.1038/nmat4369. 
            
            \bibitem{36} Cameron F Holder and Raymond E Schaak. Tutorial on Powder X-ray Diffraction for Characterizing Nanoscale Materials. \textit{ACS Nano}. (2019) \textbf{13}:7359-7365. doi:10.1021/acsnano.9b05157.
            
            \bibitem{37} Paul K Hansma. Scanning tunneling microscopy. \textit{Journal of Applied Physics}. (1987) \textbf{61}:1-24. doi:10.1063/1.338189.
            
            \bibitem{38} Feist A, Bach N, da Silva N R, Danz T, Moller M, Priebe K E, et al. Ultrafast transmission electron microscopy using a laser-driven field emitter: Femtosecond resolution with a high coherence electron beam. \textit{Ultramicroscopy}. (2017) \textbf{176}:63-73. doi:10.1016/j.ultramic.2016.12.005.
            
            \bibitem{39} Carstens T, Ispas A, Borisenko N, Atkin R, Bund A and Endres F. In situ scanning tunneling microscopy (STM), atomic force microscopy (AFM) and quartz crystal microbalance (EQCM) studies of the electrochemical deposition of tantalum in two different ionic liquids with the 1-butyl-1-methylpyrrolidinium cation. \textit{Electrochimica Acta}. (2016) \textbf{197}:374-387. doi:10.1016/j.electacta.2015.07.178.
            
            \bibitem{40} Kelly K F, Chiang I W, Mickelson E T, Hauge R H, Margrave J L, Wang X, et al. Insight into the mechanism of sidewall functionalization of single-walled nanotubes: an STM study. \textit{Chemical physics letters}. (1999) \textbf{313}:445-450. doi:10.1016/S0009-2614(99)00973-2.
            
            \bibitem{41} Maitra U, Matte H S S, Kumar P, and Rao C N R. Strategies for the synthesis of graphene, graphene nanoribbons, nanoscrolls and related materials. \textit{CHIMIA International Journal for Chemistry}. (2012) \textbf{66}:941-948. doi:10.2533/chimia.2012.941.
            
            \bibitem{42} Zhang J, Xiao J, Meng X, Monroe C, Huang Y, and Zuo J M. Free folding of suspended graphene sheets by random mechanical stimulation. \textit{Physical review letters}. (2010) \textbf{104}:166805. doi:10.1103/PhysRevLett.104.166805.
            
            \bibitem{43} Van der Lit J, Jacobse P H, Vanmaekelbergh D and Swart I. Bending and buckling of narrow armchair graphene nanoribbons via STM manipulation. \textit{New Journal of Physics}. (2015) \textbf{17}:053013. doi:10.1088/1367-2630/17/5/053013.
            
            \bibitem{44} Warner J H, Rummeli M H, Gemming T, Buchner B, Briggs G A D. Direct imaging of rotational stacking faults in few layer graphene. \textit{Nano letters}. (2008) \textbf{9}:102-106. doi:10.1021/nl8025949.
            
            \bibitem{45} Lee J K, Lee S C, Ahn J P, Kim S C, Wilson J I, John P. The growth of AA graphite on (111) diamond. \textit{The Journal of chemical physics}. (2008) \textbf{129}:234709. doi:10.1063/1.2975333.
            
            \bibitem{46} Yin C, Xiang H, Li C, Porwal H, Fang L. Low‐temperature sintering and thermal stability of Li$_2$GeO$_3$‐based microwave dielectric ceramics with low permittivity. \textit{J Am Ceram Soc}. (2018) \textbf{101}: 4608-4614. doi:10.1111/jace.15723.
            
            \bibitem{47} Li G, Luican A, Dos Santos J L, Neto A C, Reina A, Kong J, and Andrei, E Y. Observation of Van Hove singularities in twisted graphene layers. \textit{Nature Physics}. (2010) \textbf{6}:109. doi:10.1038/nphys1463.
            
            \bibitem{48} Kataura H, Kumazawa Y, Maniwa Y, Umezu I, Suzuki S, Ohtsuka Y and Achiba Y. Optical properties of single-wall carbon nanotubes. \textit{Synthetic metals}. (1999) \textbf{103}:2555-2558. doi:10.1016/S0379-6779(98)00278-1.
            
            \bibitem{49} C Rincon, S M Wasim, G Marıin, and G Snchez Perez. Optical absorption spectra near the fundamental band edge in Cu$_2$In$_4$Se$_7$Cu$_2$In$_4$Se$_7$ bulk crystals. \textit{Journal of Applied Physics}. (2003) \textbf{93}:8939-8944. doi:10.1063/1.1567800.
            
            \bibitem{50} A. N. Trukhin, U. Rogulis and M. Spingis. \textit{Journal of Luminescence}. (1997) \textbf{72-74}: 890-892. doi:10.1016/S0022-2313(96)00418-8.
            
            \bibitem{51} Donghui Lu, Inna M Vishik, Ming Yi, Yulin Chen, Rob G Moore, Zhi-Xun Shen. Angle-Resolved Photoemission Studies of Quantum Materials, \textit{Annual Review of Condensed Matter Physics}. (2012) \textbf{3}:129-167.
            doi:10.1146/annurev-conmatphys-020911-125027.
            
            \bibitem{52} Gruneis A, Attaccalite C, Pichler T, Zabolotnyy V, Shiozawa H,  Molodtsov S, Inosov D, Koitzsch A, Knupfer M, Schiessling J. Electron-electron correlation in graphite: a combined angle-resolved photoemission and first-principles study. \textit{Physical review letters}.  (2008) \textbf{100}:037601. doi:10.1103/PhysRevLett.100.037601.
            
            \bibitem{53} Ohta T, Bostwick A, McChesney J L, Seyller T, Horn K, Rotenberg E. Interlayer interaction and electronic screening in multilayer graphene investigated with angle-resolved photoemission spectroscopy. \textit{Physical Review Letters}. (2007) \textbf{98}:206802. doi:10.1103/PhysRevLett.98.206802.
            
            \bibitem{54} Kim K, Walter A, Moreschini L, et al. Coexisting massive and massless Dirac fermions in symmetry-broken bilayer graphene. \textit{Nature Mater}. (2013) \textbf{12}:887-892. doi:10.1038/nmat3717.
            
            
            \bibitem{55} Coletti C, Forti S, Principi A, Emtsev K V, Zakharov A A, Daniels K M, Daas B K, Chandrashekhar M, Ouisse T, Chaussende D. Revealing the electronic band structure of trilayer graphene on SiC: An angle-resolved photoemission study. \textit{Physical Review B}. (2013) \textbf{88}:155439. doi:10.1103/PhysRevB.88.155439.
            
            \bibitem{56} R Balog, B Jrgensen, L Nilsson, M Andersen, E Rienks, M Bianchi, M Fanetti, E Lgsgaard, A Baraldi, S Lizzit, Z Sljivancanin, F Besenbacher, B Hammer, T G Pedersen, P Hofmann and L Horneker. Bandgap opening in graphene induced by patterned hydrogen adsorption. \textit{Nat. Mater}. (2010) \textbf{9}: 315-319. doi:10.1038/nmat2710.
            
            \bibitem{57} Y Niimi, K Kanisawa, H Kojima, H Kambara, Y Hirayama, S Tarucha and Hiroshi Fukuyama1. STM/STS Measurements of Two-Dimensional Electronic States in Magnetic Fields at Epitaxially Grown InAs(111)A Surfaces. \textit{Journal of Physics: Conference Series}. (2007) \textbf{61}:174. doi:10.1088/1742-6596/61/1/174.
            
            \bibitem{58} Huang H, Wei D, Sun J, Wong S L, Feng Y P, Neto A H C, Wee A T S. Spatially resolved electronic structures of atomically precise armchair graphene nanoribbons. \textit{Scientific Reports}. (2012) \textbf{2}:983. doi:10.1038/srep00983.
                        
            \bibitem{59} Odom T W, Huang J L, Kim P Lieber C M. Atomic structure and electronic properties of single-walled carbon nanotubes. \textit{Nature} (1998) \textit{391}:62-64. doi:10.1038/34145.
                        
            \bibitem{60} Klusek Z. Investigations of splitting of the $\pi$ bands in graphite by scanning tunneling spectroscopy. \textit{Applied Surface Science}. (1999) \textbf{151}:251-261. doi:10.1016/S0169-4332(99)00282-2.
            
            \bibitem{61} Luican A, Li G, Reina A, Kong J, Nair R, Novoselov K S, Geim A K, Andrei E. Single-layer behavior and its breakdown in twisted graphene layers. \textit{Physical Review Letters}. (2011) \textbf{106}:126802. doi:10.1103/PhysRevLett.106.126802.
            
            \bibitem{62} Li G, Luican A, Dos Santos J L, Neto A C, Reina A Kong J, Andrei E. Observation of Van Hove singularities in twisted graphene layers. \textit{Nature Phys}. (2010) \textbf{6}:109-113. doi:10.1038/nphys1463.
            
            \end{thebibliography}
\end{document}